\documentclass{aa}
\usepackage{graphics}

\begin{document}

\title{Iron abundance in \ion{H}{ii}~regions
\thanks{Based on observations made with the Isaac Newton Telescope, operated
on the island of La Palma by the Isaac Newton Group in the Spanish
Observatorio del Roque de los Muchachos of the Instituto de Astrof\'\i sica
de Canarias.}}

\author{M. Rodr\'\i guez}

\institute{Instituto de Astrof\'\i sica de Canarias, E-38200 La Laguna, 
Tenerife, Canarias, Spain
\thanks{{\emph Present address}: Instituto Nacional de Astrof\'\i sica, \'Optica
y Electr\'onica INAOE, Apdo Postal 51 y 216, 72000 Puebla, Pue., M\'exico
(mrodri@inaoep.mx)}}

\date{Received date; accepted date}

\abstract{
Optical CCD spectra are used to determine the \element{Fe} abundances at
several positions inside seven bright Galactic \ion{H}{ii} regions.
The observed [\ion{Fe}{iii}] line ratios are compared with the predictions of
different sets of collision strengths and transition probabilities for this ion
to select the atomic data providing the best fit to the observations.
The values found for the \element[++]{Fe} and \element[+]{Fe} abundances,
along with ionization correction factors for the contribution of
\element[3+]{Fe}, obtained from available grids of photoionized models, 
imply that the Fe/O ratio in the ionized gas is between 2\% and 30\% of
solar.
The \element{Fe} abundances derived for each area are correlated both with the
degree of ionization and the colour excess.
A possible explanation is suggested, namely the presence of a population of
small grains, probably originating from the fragmentation of larger grains.
These small grains would release \element{Fe} atoms into the gas after the
absorption of energetic photons; the small grains surviving this destruction
process would be swept out of the ionized region by the action of radiation
pressure or stellar winds.
An indication of a further and more efficient destruction agent is given
by the high \element{Fe} abundance derived for a position sampling the optical
jet \object{H~399} in \object{M20}, where dust destruction due to shock waves
has presumably taken place.
}

\maketitle
\section{Introduction}

The first reliable measurement of the \element{Fe} abundance in an \ion{H}{ii}
region was carried out for \object{M42} by Osterbrock et al.\ (\cite{otv92}),
who found $\element{Fe}/\element{H}$ to be underabundant by a factor of 12 with
respect to the solar value, confirming the previous estimate by Olthof \&
Pottasch (\cite{olt75}).
Similar results were obtained for \object{M17} and \object{M8}, for which
Peimbert et al.\ (\cite{pei93}) found underabundance factors of 4 and 20,
respectively.
These values are consistent with the expected depletion of \element{Fe} on to
the dust grains known to coexist with the ionized gas
(M\"unch \& Persson \cite{mun71}; Osterbrock \cite{ost89}), but the derived
\element{Fe} abundances are significantly above the typical values for the
undisturbed dense interstellar medium, $\sim100$ below solar (Jenkins
\cite{jen87}).
This difference in the \element{Fe} depletion factors in \ion{H}{ii} regions
and the interstellar medium is not readily explained, since refractory dust
grains are expected to survive inside \ion{H}{ii} regions (Osterbrock
\cite{ost89}).

A possible explanation was offered in Rodr\'\i guez (\cite{rod96}), where the
gaseous \element{Fe} abundances at different positions within seven Galactic
\ion{H}{ii} regions were shown to be correlated with the degree of ionization, and
this correlation was interpreted by relating the energetic photons with the
release of \element{Fe} atoms from dust grains.
This paper presents a more detailed and complete analysis of the data used in
Rodr\'\i guez (\cite{rod96}) to determine the \element{Fe} abundances in
\object{M42}, \object{M43}, \object{M8}, \object{M16}, \object{M17},
\object{M20} and \object{NGC 7635}.
The present analysis relies on recent atomic data for the determination of
physical conditions and ionic abundances, and uses a somewhat different
approach to derive the \element{Fe} abundance, leading to abundance values
lower by about 40\% than those presented in Rodr\'\i guez (\cite{rod96}).
The correlation of the \element{Fe} abundance with the degree of ionization is
confirmed to hold, as well as the further correlation with the colour excess,
also previously presented in Rodr\'\i guez (\cite{rod96}).
The implications that both results have for dust in \ion{H}{ii} regions are
discussed.

\section{The data}

The results in this paper are based on the line intensities measured by
Rodr\'\i guez (\cite{rod96}, \cite{rod99b}) in the
$\lambda\lambda4200\mbox{--}8800$ spectra of several positions within seven
Galactic \ion{H}{ii} regions: \object{M42}, \object{M43}, \object{M8},
\object{M16}, \object{M17}, \object{M20} and \object{NGC 7635}.
This work centres on the \element{Fe} lines, but the values derived by
Rodr\'\i guez (\cite{rod99b}) for some physical conditions and
ionic abundances are also used: $N_{\rm e}[\ion{S}{ii}]$, the densities implied
by the intensity ratio of [\ion{S}{ii}] lines $I(\lambda6716)/I(\lambda6731)$;
$T_{\rm e}[\ion{N}{ii}]$, the temperatures derived from the intensity ratio of
[\ion{N}{ii}] lines $I(\lambda6548+\lambda6583)/I(\lambda5754)$; and the
ionic abundance ratios $\element[+]{N}/\element[+]{H}$,
$\element[+]{O}/\element[+]{H}$,
$\element[++]{O}/\element[+]{H}$ and $\element{O}/\element{H}=
\element[+]{O}/\element[+]{H}+\element[++]{O}/\element[+]{H}$.

The ionization potentials for \element[0]{Fe}, \element[+]{Fe},
\element[++]{Fe} and \element[3+]{Fe} are 7.9, 16.2, 30.6 and 54.8~eV,
respectively.
Considering the excitation characteristics of the objects in the sample and the
absence of \ion{He}{ii}~$\lambda$4686 emission from all the nebular spectra
(the ionization potential for \element[+]{He} is 54.4~eV)
\element[3+]{Fe} is not expected to be further ionized, and only
\element[+]{Fe}, \element[++]{Fe} and \element[3+]{Fe} should have appreciable
concentrations within the ionized gas.
Several [\ion{Fe}{iii}] and [\ion{Fe}{ii}] lines can be measured in the optical
spectra of \ion{H}{ii} regions, but no optical [\ion{Fe}{iv}] lines have been
detected in \ion{H}{ii} regions and only one line has been measured in the UV
spectra of an \ion{H}{ii} region: [\ion{Fe}{iv}]~$\lambda2837$ in \object{M42}
(Rubin et al.\ \cite{rub97}).
Therefore, the values derived here for the \element{Fe} abundance are based on
the $\element[+]{Fe}$ and $\element[++]{Fe}$ abundances and ionization
correction factors for the contribution of \element[3+]{Fe} to the total
abundance derived from available photoionization models.

\section{[\ion{Fe}{iii}] lines}

Up to twelve [\ion{Fe}{iii}] lines, previously listed and measured in
\object{M42} by Osterbrock et al.\ (\cite{otv92}) and belonging to
($1F$), ($2F$) and ($3F$) multiplets, were detected in each object.
At the achieved spectral resolution of 4~\AA, [\ion{Fe}{iii}]~$\lambda$4986
and [\ion{Fe}{iii}]~$\lambda$4987 are blended;
[\ion{Fe}{iii}]~$\lambda$4607 is contaminated by \ion{O}{ii}~$\lambda$4609 and
\ion{N}{ii}~$\lambda$4607 (Esteban et al.\ \cite{est98}); the intensity
measured for [\ion{Fe}{iii}]~$\lambda$4778 might have a small contribution from
\ion{N}{ii}~$\lambda$4780 (Esteban et al.\ \cite{est99a}), and
[\ion{Fe}{iii}]~$\lambda$4658 and [\ion{Fe}{iii}]~$\lambda$5270 are appreciably
blended in some objects with \ion{O}{ii}~$\lambda$4661 and
[\ion{Fe}{ii}]~$\lambda$5273, respectively.
On the basis of these considerations, [\ion{Fe}{iii}]~$\lambda$4607 has been
excluded from the abundance determination, [\ion{Fe}{iii}]~$\lambda$4778 has
 been considered only in those positions where the abundance implied by this
feature is consistent with the mean abundance implied by the other lines, and
[\ion{Fe}{iii}]~$\lambda$4658 and [\ion{Fe}{iii}]~$\lambda$5270 have  been
used only 
for those objects where the contamination of these lines is negligible.

\begin{table*}
\caption[ ]{[\ion{Fe}{iii}] line intensities and \element[++]{Fe} abundance}
\begin{tabular}{llllllllllll}
\hline
\noalign{\smallskip}
\multicolumn{1}{l}{Object}&
     \multicolumn{10}{c}{$I(\lambda)/I(\mbox{H}\beta)\times 100^{\rm a}$} &
     \multicolumn{1}{l}{$\element[++]{Fe}/\element[+]{H}\pm\,\sigma$}\\
\noalign{\smallskip}
\cline{2-11}
\noalign{\smallskip}
& $\lambda$4658 & $\lambda$4702 & $\lambda$4734 & $\lambda$4755 &
	 $\lambda$4769 & $\lambda$4778 & $\lambda$4881 & $\lambda$4986+7 &
	 $\lambda$5270 & $\lambda$5412 & \\
\noalign{\smallskip}
\hline
\noalign{\smallskip}
\object{M42}~A--1	& 0.77  & 0.255 & 0.096 & 0.142 & 0.086 &
			  0.081:& 0.32  & 0.056 & blend & 0.029 &
			  (3.3$\pm$0.4)$\times10^{-7}$\\
\object{M42}~A--2	& 0.69  & 0.208 & 0.082 & 0.132 & 0.071 &
			  0.061 & 0.263 &$\dots$& blend & 0.026 &
			  (2.2$\pm$0.4)$\times10^{-7}$\\
\object{M42}~A--3	& 0.64  & 0.189 & 0.065 & 0.113 & 0.065 &
			  0.029 & 0.255 &$\dots$& blend & 0.015 &
			  (2.5$\pm$0.4)$\times10^{-7}$\\
\object{M42}~A--4	& 0.72  & 0.23  & 0.090 & 0.153 & 0.074 &
			  0.064:& 0.289 &$\dots$& blend & 0.021 &
			  (2.7$\pm$0.4)$\times10^{-7}$\\
\object{M42}~A--5	& 0.72  & 0.20  & 0.074 & 0.132 & 0.047 &
			  0.042 & 0.25  &$\dots$& blend & 0.015:&
			  (3.6$\pm$0.6)$\times10^{-7}$\\
\object{M42}~A--6	& 0.685 & 0.21  & 0.06  & 0.14  &$\dots$&
			 $\dots$& 0.249 &$\dots$& blend &$\dots$&
			  (2.5$\pm$0.2)$\times10^{-7}$\\
\noalign{\smallskip}
\hline
\noalign{\smallskip}
\object{M42}~B--1	& 0.87  & 0.251 & 0.088 & 0.179 & 0.081 &
			  0.052 & 0.352 & 0.065 & blend & 0.031 &
			  (4.2$\pm$0.7)$\times10^{-7}$\\
\object{M42}~B--2	& 0.905 & 0.248 & 0.110 & 0.186 & 0.094 &
			  0.047 & 0.364 & 0.05: & blend & 0.023 &
			  (4.7$\pm$0.8)$\times10^{-7}$\\
\object{M42}~B--3	& 0.85  & 0.219 & 0.067 & 0.18  & 0.074 &
			  0.052 & 0.309 & 0.053:& blend & 0.018 &
			  (4.2$\pm$1.0)$\times10^{-7}$\\
\object{M42}~B--4	& 1.34  & 0.384 & 0.149 & 0.262 & 0.128 &
			  0.083 & 0.518 & 0.118 & blend & 0.053 &
			  (6.7$\pm$1.0)$\times10^{-7}$\\
\object{M42}~B--5	& 0.894 & 0.25  & 0.07  & 0.14  & 0.08  &
			  0.05  & 0.287 & 0.11  & blend &$\dots$&
			  (4.8$\pm$1.1)$\times10^{-7}$\\
\object{M42}~B--6	& 0.90  & 0.26  & 0.08  & 0.146 & 0.10  &
			  0.06  & 0.33  & 0.08  & blend & 0.038 &
			  (4.9$\pm$0.7)$\times10^{-7}$\\
\noalign{\smallskip}
\hline
\noalign{\smallskip}
\object{M43}--1		& 0.8   & 0.16  &$\dots$& 0.19  &$\dots$&
			 $\dots$& 0.24  & 0.178:& blend &$\dots$&
			 (8.6$\pm$1.9)$\times10^{-7}$\\
\object{M43}--2		& 0.81  & 0.23  &$\dots$& 0.33: &$\dots$&
			 $\dots$& 0.249 & 0.18: & blend &$\dots$&
			 (8.6$\pm$0.9)$\times10^{-7}$\\
\object{M43}--3		& 0.6   &$\dots$&$\dots$& 0.12  &$\dots$&
			 $\dots$&$\dots$& 0.21  & blend &$\dots$&
			 (5.5$\pm$0.9)$\times10^{-7}$\\
\object{M43}--4		& 0.83  & 0.21  &$\dots$& 0.19  &$\dots$&
			 $\dots$& 0.209 & 0.164:& blend &$\dots$&
			 (9.4$\pm$1.2)$\times10^{-7}$\\
\object{M43}--5		& 0.94  & 0.18  &$\dots$& 0.19  &$\dots$&
			 $\dots$& 0.32  & 0.17: & blend &$\dots$&
			 (1.0$\pm$0.2)$\times10^{-6}$\\
\noalign{\smallskip}
\hline
\noalign{\smallskip}
\object{M8}--1		& 0.57  & 0.18  & 0.07  & 0.11  & 0.06  &
			 $\dots$& 0.25  & 0.07  & blend &$\dots$&
			 (3.2$\pm$0.4)$\times10^{-7}$\\
\object{M8}--2		& 0.57  & 0.18  & 0.05  & 0.065 &$\dots$&
			 $\dots$& 0.21  & 0.06  & blend &$\dots$&
			 (3.1$\pm$0.9)$\times10^{-7}$\\
\object{M8}--3		& 0.64  & 0.19  &$\dots$& 0.13  & 0.04  &
			 $\dots$& 0.23  &$\dots$& blend &$\dots$&
			 (3.7$\pm$0.7)$\times10^{-7}$\\
\object{M8}--4		& 0.55  & 0.19  & 0.05  & 0.10  & 0.037 &
			  0.02  & 0.178 & 0.052 & blend &$\dots$&
			 (3.2$\pm$0.8)$\times10^{-7}$\\
\object{M8}--5		& 0.51  & 0.23: &$\dots$& 0.11  &$\dots$&
			 $\dots$& 0.15  & 0.09  & blend &$\dots$&
			 (3.5$\pm$1.1)$\times10^{-7}$\\
\object{M8}--6		& 0.57  & 0.11  &$\dots$&$\dots$&$\dots$&
			 $\dots$& 0.11  & 0.10  & blend &$\dots$&
			 (2.3$\pm$0.5)$\times10^{-7}$\\
\noalign{\smallskip}
\hline
\noalign{\smallskip}
\object{M16}--1		& 0.17  &$\dots$&$\dots$& 0.06  &$\dots$&
			 $\dots$& 0.044 &$\dots$& blend &$\dots$&
			 (1.5$\pm$0.8)$\times10^{-7}$\\
\object{M16}--2		&$\dots$&$\dots$&$\dots$&$\dots$&$\dots$&
			 $\dots$&$\dots$& 0.31  &$\dots$&$\dots$&
			 2.0$\times10^{-7}$\\
\noalign{\smallskip}
\hline
\noalign{\smallskip}
\object{M17}--1		& blend &$\dots$&$\dots$&$\dots$&$\dots$&
			 $\dots$& 0.039 &$\dots$& 0.125 &$\dots$&
			 (1.2$\pm$0.5)$\times10^{-7}$\\
\object{M17}--2		& blend &$\dots$&$\dots$& 0.08  &$\dots$&
			 $\dots$& 0.07  &$\dots$& 0.12  &$\dots$&
			 (1.5$\pm$0.5)$\times10^{-7}$\\
\object{M17}--3		& blend &$\dots$&$\dots$&$\dots$&$\dots$&
			 $\dots$& 0.07  &$\dots$& 0.095 &$\dots$&
			 (1.3$\pm$0.3)$\times10^{-7}$\\
\noalign{\smallskip}
\hline
\noalign{\smallskip}
\object{M20}--1		& 1.0   & 0.18  &$\dots$& 0.19  &$\dots$&
			 $\dots$& 0.30  & 0.29: & blend &$\dots$&
			 (7.8$\pm$1.9)$\times10^{-7}$\\
\object{M20}--2		& 0.3   &$\dots$&$\dots$&$\dots$&$\dots$&
			 $\dots$& 0.04  & 0.23  & blend &$\dots$&
			 (2.2$\pm$0.6)$\times10^{-7}$\\
\object{M20}--3		& 0.4   &$\dots$&$\dots$&$\dots$&$\dots$&
			 $\dots$&$\dots$&$\dots$&$\dots$&$\dots$&
			 2.9$\times10^{-7}$\\
\noalign{\smallskip}
\hline
\noalign{\smallskip}
\object{NGC~7635}--1	& 0.14  & 0.06  &$\dots$&$\dots$&$\dots$&
			 $\dots$& 0.122 & 0.04  & blend &$\dots$&
			 (1.7$\pm$0.6)$\times10^{-7}$\\
\object{NGC~7635}--2	& 0.39  & 0.11  &$\dots$& 0.08  &$\dots$&
			 $\dots$& 0.18  & 0.046 & blend &$\dots$&
			 (2.5$\pm$0.2)$\times10^{-7}$\\
\object{NGC~7635}--3	& 0.38  &$\dots$&$\dots$& 0.07  &$\dots$&
			 $\dots$& 0.11  & 0.04  & blend &$\dots$&
			 (2.1$\pm$0.3)$\times10^{-7}$\\
\noalign{\smallskip}
\hline
\noalign{\smallskip}
\multicolumn{12}{l}{$^{\rm a}$ [\ion{Fe}{iii}]~$\lambda$4658 and
	[\ion{Fe}{iii}]~$\lambda$5271 are blended
	with \ion{O}{ii}~$\lambda$4661 and [\ion{Fe}{ii}]~$\lambda$5273,
	respectively. The line intensities marked}\\
\multicolumn{12}{l}{with colons imply abundance values that deviate by more
	than 50\% from the mean value derived from the other lines}\\
\multicolumn{12}{l}{and have been excluded from the abundance determination}\\
\end{tabular}
\end{table*}

The reddening-corrected intensity ratios of the [\ion{Fe}{iii}] lines relative
to \element{H}$\beta$ are listed in Table~1 for all the positions studied (see
Rodr\'\i guez \cite{rod99b} for the values of the parameters used in the
extinction correction).
The extinction corrections applied to the relative [\ion{Fe}{iii}] line
intensities are very small, a few per cent in most cases, with a maximum of
30\% for [\ion{Fe}{iii}]~$\lambda$5270 in \object{M17}.
The accuracy of the line intensities in Table~1 may vary from 10--20\% for the
positions in \object{M42} to more than 50\% for some cases, in particular for
those positions where only one [\ion{Fe}{iii}] line could be measured.

\section{Atomic data for [\ion{Fe}{iii}]}

\begin{figure*}
  \resizebox{\hsize}{!}{\includegraphics{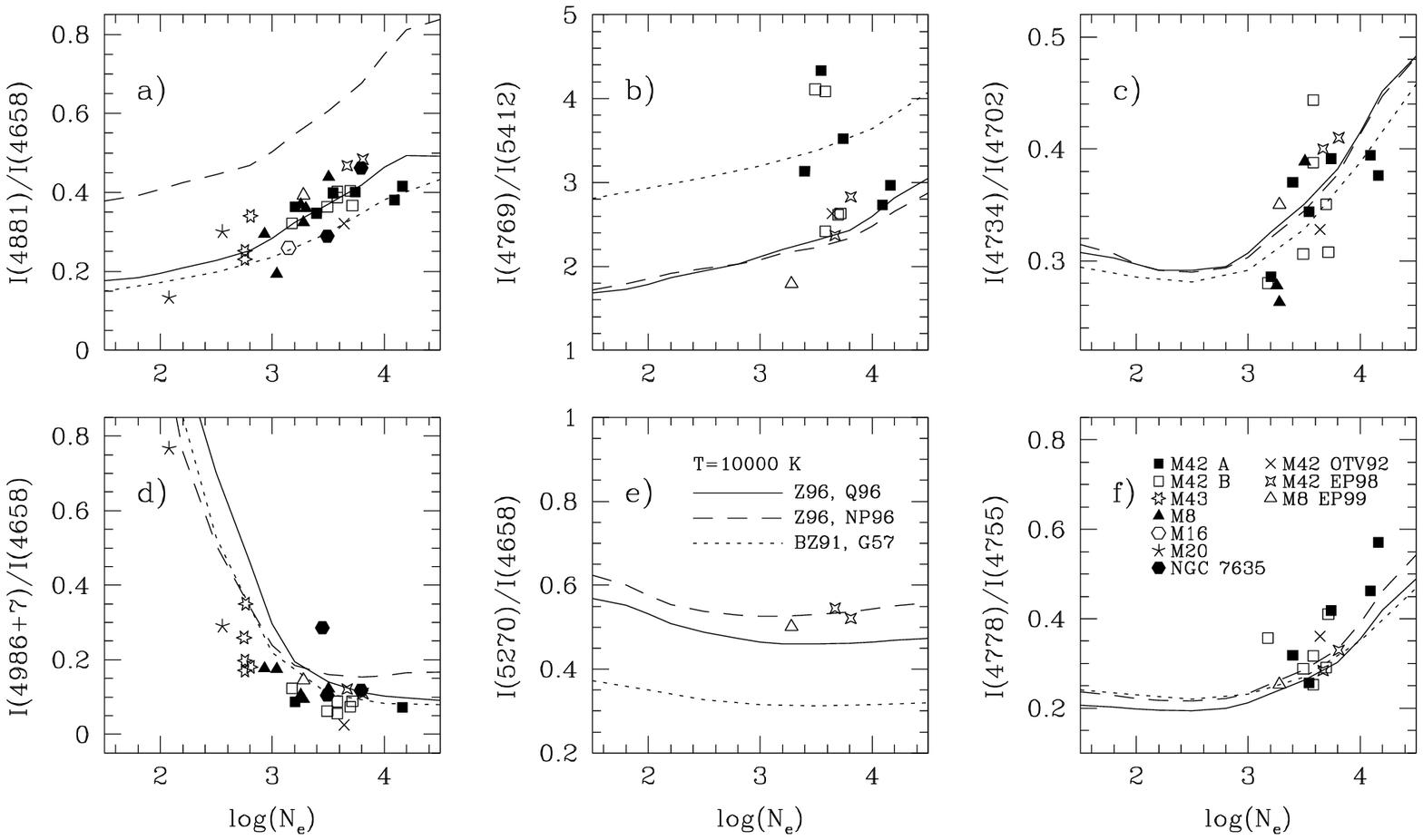}}
    \caption{Several observed [\ion{Fe}{iii}] line ratios are
    presented as a function of density, $N_{\rm e}[\ion{S}{ii}]$, and compared
    with the collisional excitation predictions for $T_{\rm e}=10\,000$~K.
    The atomic data used in the calculations are identified in panel {\bf e},
    where Z96, Q96, NP96, BZ91 and G57 stand for Zhang (\cite{zha96}), Quinet
    (\cite{qui96}), Nahar \& Pradhan (\cite{nah96}), Berrington et al.\
    (\cite{ber91}) and Garstang (\cite{gar57}), respectively. The symbols
    representing the different \ion{H}{ii} regions are identified in panel
    {\bf f}, where OTV92, EP98 and EP99 identify the measurements performed by
    Osterbrock et al.\ (\cite{otv92}) and Esteban et al.\ (\cite{est98},
    \cite{est99a})}
    \label{fig1}
\end{figure*}

Keenan et al.\ (\cite{kee92}) derived the relative populations of the lowest 17
levels of the $3d^6$ configuration in \element[++]{Fe} using the
collision strengths of Berrington et al.\ (\cite{ber91}) and the
transition probabilities of Garstang (\cite{gar57}).
Further calculations of the collision strengths for this ion are reported
by Zhang (\cite{zha96}), whereas new values for the transition probabilities
for the forbidden transitions have been derived by Nahar \& Pradhan
(\cite{nah96}) and Quinet (\cite{qui96}).
The new effective collision strengths, when compared with the previous values
of Berrington et al.\ (\cite{ber91}), are enhanced by a factor of up to 2,
whereas the three available sets of transition probabilities differ from each
other by $\sim30$\% for most transitions, with differences up to an order of
magnitude or more for some transitions.

In order to compare the results implied by the most recent atomic data with the
previous calculations of Keenan et al.\ (\cite{kee92}),
the level populations for \element[++]{Fe} have been calculated with the
collision strengths reported by Zhang (\cite{zha96}) and the two recent sets of
transition probabilities (Nahar \& Pradhan \cite{nah96}; Quinet \cite{qui96}).
All collisional and downward radiative transitions for the 34 levels of the
$3d^6$ configuration have been considered; test calculations restricted to the
lowest 17 levels considered by Keenan et al.\ (\cite{kee92}) are found to agree
to within 10\% with the 34-level calculations.

In Fig.~\ref{fig1} some observed and predicted [\ion{Fe}{iii}] line ratios are
presented as a function of $N_{\rm e}[\ion{S}{ii}]$ (Rodr\'\i guez
\cite{rod99b}).
The values found for $T_{\rm e}[\ion{N}{ii}]$ cover the range
$T_{\rm e}[\ion{N}{ii}]=7800\mbox{--}10\,800$~K (Rodr\'\i guez \cite{rod99b}),
but the predicted line ratios for any temperature in this range are similar,
and only the calculations for $T_{\rm e}=10\,000$~K are shown in
Fig.~\ref{fig1}.
The [\ion{Fe}{iii}] line ratios measured by Osterbrock et al.\ (\cite{otv92})
and Esteban et al.\ (\cite{est98}, \cite{est99a}) in \object{M42} and
\object{M8} are also presented in this figure; note that the measurements of
Esteban et al.\ (\cite{est98}, \cite{est99a}), performed at high-resolution,
can be considered the most reliable.

Figures~\ref{fig1}a, \ref{fig1}b, \ref{fig1}d and \ref{fig1}e show some of the
line ratios for which the different atomic parameters lead to the greatest
differences.
In general, all the atomic data fit the measured line ratios reasonably well,
but the results in Figs.~\ref{fig1}a and \ref{fig1}e suggest that the
calculations based on the collision strengths of Zhang (\cite{zha96}) and the
transition probabilities of Quinet (\cite{qui96}) lead to a better fit.

The best way to judge the accuracy of the atomic data is to consider the
dispersion in the abundances implied by the different lines.
The most accurate [\ion{Fe}{iii}] intensities are probably those measured by
Esteban et al.\ (\cite{est98}) in \object{M42}.
The abundances and dispersions implied by the intensities measured in
their position~1 for the ten lines in Table~1 and the three combinations of
atomic data considered here are the following:
$\element[++]{Fe}/\element[+]{H}=(3.5\pm0.8)\times10^{-7}$ (Keenan et al.\
\cite{kee92}),
$(2.5\pm0.3)\times10^{-7}$ (Zhang \cite{zha96}; Nahar \& Pradhan \cite{nah96})
and $(2.5\pm0.2)\times10^{-7}$ (Zhang \cite{zha96}; Quinet \cite{qui96}).
When the [\ion{Fe}{iii}] intensities measured in other regions are considered,
similar results are obtained: the oldest atomic data lead in most cases to
higher dispersions and 10 to 50\% higher mean abundance values. The most
recent collision strengths lead to essentially equal mean abundance values
independently of the transition probabilities used, and the dispersions are
similar for the two recent sets of transition probabilities, although the
transition probabilities of Quinet (\cite{qui96}) lead in general to somewhat
lower dispersions.
Therefore, the collision strengths of Zhang (\cite{zha96}) and the transition
probabilities of Quinet (\cite{qui96}) will be used hereafter to derive the
\element[++]{Fe} abundances.

\section{$\element[++]{Fe}/\element[+]{H}$}

The last column in Table~1 shows the values of
$\element[++]{Fe}/\element[+]{H}$ derived at each position from the measured
[\ion{Fe}{iii}] intensities by using the collision strengths of Zhang
(\cite{zha96}), the transition probabilities of Quinet (\cite{qui96}),
the \ion{H}{i} emissivities of Hummer \& Storey (\cite{hum87}) and the physical
conditions $T_{\rm e}[\ion{N}{ii}]$, $N_{\rm e}[\ion{S}{ii}]$
(Rodr\'\i guez \cite{rod99b}).
The dispersions in the abundances implied by the different lines considered are
also indicated in Table~1; it can be seen that most of them are quite low, in
the range 10--20\%.

The values derived for $\element[++]{Fe}/\element[+]{H}$ are lower than those
presented in Rodr\'\i guez (\cite{rod96}) by 10--70\%.
Up to 20\% in these differences is due to the new values of the physical
conditions implied by updated atomic data (Rodr\'\i guez \cite{rod99b});
the remaining differences arise from the use of the most recent atomic data for
\element[++]{Fe}.

\section{[\ion{Fe}{ii}] emission and \element[+]{Fe} abundance}

Many lines of [\ion{Fe}{ii}] have been identified in the optical spectra of
\ion{H}{ii} regions (see, for example, Osterbrock et al.\ \cite{otv92};
Rodr\'\i guez \cite{rod96}; Esteban et al.\ \cite{est98}, \cite{est99a}).
Most of these lines are severely affected by fluorescence effects
(Rodr\'\i guez \cite{rod99a}), but an estimate of the contribution of
\element[+]{Fe} to the total \element{Fe} abundance can be obtained from the
intensity measured for [\ion{Fe}{ii}]~$\lambda$8617, which is almost
insensitive to the effects of UV pumping (Lucy \cite{lucy95}; Baldwin et al.\
\cite{bald96}).
The reddening-corrected intensities of [\ion{Fe}{ii}]~$\lambda$8617 relative
to Pa12 or Pa13 are listed in Table~2; the applied extinction
corrections are below 10\%, and the uncertainties in the relative intensities
may vary widely from $\sim$15\% in \object{M42} to more than 50\% in
\object{M20}--2, \object{M16} and \object{M17}.
The values of $\element[+]{Fe}/\element[+]{H}$ have been derived from the
intensities in Table~2 by using the emissivities for [\ion{Fe}{ii}] (Bautista
\& Pradhan \cite{bp96}) and \ion{H}{i} (Hummer \& Storey \cite{hum87})
interpolated to the physical conditions $T_{\rm e}[\ion{N}{ii}]$ and
$N_{\rm e}[\ion{S}{ii}]$.

\begin{table}
\caption[ ]{[\ion{Fe}{ii}]~$\lambda$8617 intensities and
	\element[+]{Fe} abundance}
\begin{tabular}{llllc}
\hline
\noalign{\smallskip}
Object & $I/I(\ion{H}{i})$ & \ion{H}{i} &
	 $\element[+]{Fe}/\element[+]{H}$ &
	 $\element[+]{Fe}/\element[++]{Fe}$ \\
\noalign{\smallskip}
\hline
\noalign{\smallskip}
\object{M42} A--1   & 0.094 & Pa12 & 7.8$\times10^{-8}$ & 0.24 \\
\object{M42} A--2   & 0.09  & Pa12 & 5.9$\times10^{-8}$ & 0.27 \\
\object{M42} A--3   & 0.030 & Pa12 & 2.4$\times10^{-8}$ & 0.10 \\
\object{M42} A--4   & 0.061 & Pa12 & 4.5$\times10^{-8}$ & 0.17 \\
\object{M42} A--5   & 0.048 & Pa12 & 4.5$\times10^{-8}$ & 0.13 \\
\object{M42} A--6   & 0.038 & Pa12 & 2.9$\times10^{-8}$ & 0.12 \\
\noalign{\smallskip}
\hline
\noalign{\smallskip}
\object{M42} B--1   & 0.088 & Pa12 & 7.9$\times10^{-8}$ & 0.19 \\
\object{M42} B--2   & 0.044 & Pa12 & 4.1$\times10^{-8}$ & 0.09 \\
\object{M42} B--3   & 0.055 & Pa12 & 5.2$\times10^{-8}$ & 0.12 \\
\object{M42} B--4   & 0.159 & Pa12 & 1.4$\times10^{-7}$ & 0.21 \\
\object{M42} B--5   & 0.08  & Pa12 & 8.5$\times10^{-8}$ & 0.18 \\
\object{M42} B--6   & 0.101 & Pa12 & 9.8$\times10^{-8}$ & 0.20 \\
\noalign{\smallskip}
\hline
\noalign{\smallskip}
\object{M43}--1     & 0.095 & Pa12 & 1.7$\times10^{-7}$ & 0.20 \\
\object{M43}--2     & 0.157 & Pa12 & 2.7$\times10^{-7}$ & 0.31 \\
\object{M43}--3     & 0.21  & Pa12 & 3.6$\times10^{-7}$ & 0.65 \\
\object{M43}--4     & 0.085 & Pa12 & 1.6$\times10^{-7}$ & 0.17 \\
\object{M43}--5     & 0.08  & Pa12 & 1.4$\times10^{-7}$ & 0.14 \\
\noalign{\smallskip}
\hline
\noalign{\smallskip}
\object{M8}--1      & 0.021 & Pa12 & 2.0$\times10^{-8}$ & 0.06 \\
\object{M8}--2      & 0.027 & Pa12 & 3.0$\times10^{-8}$ & 0.10 \\
\object{M8}--3      & 0.024 & Pa12 & 2.5$\times10^{-8}$ & 0.07 \\
\object{M8}--4      & 0.013 & Pa12 & 1.5$\times10^{-8}$ & 0.05 \\
\object{M8}--5      & 0.019 & Pa12 & 2.6$\times10^{-8}$ & 0.07 \\
\object{M8}--6      & 0.02  & Pa12 & 2.2$\times10^{-8}$ & 0.10 \\
\noalign{\smallskip}
\hline
\noalign{\smallskip}
\object{M16}--1     & 0.015 & Pa12 & 1.9$\times10^{-8}$ & 0.13 \\
\object{M16}--2     &$\dots$&$\dots$& $\dots$ & $\dots$ \\
\noalign{\smallskip}
\hline
\noalign{\smallskip}
\object{M17}--1     & 0.009 & Pa12 & 1.1$\times10^{-8}$ & 0.09 \\
\object{M17}--2     &$\dots$&$\dots$& $\dots$ & $\dots$ \\
\object{M17}--3     & 0.016 & Pa13 & 1.5$\times10^{-8}$ & 0.12 \\
\noalign{\smallskip}
\hline
\noalign{\smallskip}
\object{M20}--1     & 0.5   & Pa12 & 7.7$\times10^{-7}$ & 0.99 \\
\object{M20}--2     & 0.1   & Pa12 & 1.9$\times10^{-7}$ & 0.86 \\
\object{M20}--3     &$\dots$&$\dots$& $\dots$ & $\dots$ \\
\noalign{\smallskip}
\hline
\noalign{\smallskip}
\object{NGC~7635}--1& 0.051 & Pa12 & 5.9$\times10^{-8}$ & 0.35 \\
\object{NGC~7635}--2& 0.066 & Pa12 & 6.7$\times10^{-8}$ & 0.27 \\
\object{NGC~7635}--3& 0.06  & Pa13 & 5.1$\times10^{-8}$ & 0.24 \\
\noalign{\smallskip}
\hline
\end{tabular}
\end{table}

In general, the values of $\element[+]{Fe}/\element[+]{H}$ are much lower than
those found for $\element[++]{Fe}/\element[+]{H}$, as can be
expected from the low ionization potential of \element[+]{Fe} (16.2~eV):
most of the positions show $\element[+]{Fe}/\element[++]{Fe}\la0.3$ (see
Table~2).
One exception is \object{M43}--3, where the $\element[++]{Fe}/\element[+]{H}$
abundance ratio is below those derived for the other positions in \object{M43},
and the value of
$\element[++]{Fe}/\element[+]{H}+\element[+]{Fe}/\element[+]{H}$
turns out to be very similar.
The most notable exceptions are \object{M20}--1 and \object{M20}--2, where
$\element[+]{Fe}/\element[++]{Fe}\sim1$.
The spectra obtained for \object{M20} are very noisy in the wavelength range
around [\ion{Fe}{ii}]~$\lambda$8617 and the intensity measured for this
line is uncertain by a factor of 2 in \object{M20}--2, but in
position \object{M20}--1 all the [\ion{Fe}{ii}] and [\ion{Fe}{iii}] lines
are relatively strong and the intensity of [\ion{Fe}{ii}]~$\lambda$8617 has a
much lower uncertainty.
Position \object{M20}--1 samples the optical jet \object{HH~399}, where the
conditions could differ from those found in other regions (see the discussion
in Sect.~7.1).

It has been suggested (Bautista \& Pradhan \cite{bp98}, and references therein)
that most of the [\ion{Fe}{ii}] emission in \object{M42} and other \ion{H}{ii}
regions originates in partially ionized layers with densities
$N_{\rm e}\sim10^6\mbox{ cm}^{-3}$, much higher than those measured
in the fully ionized zones, where $N_{\rm e}\le10^4\mbox{ cm}^{-3}$, but
several arguments have been presented against this hypothesis (Baldwin et
al.\ \cite{bald96}; Esteban et al.\ \cite{est99b}; Rodr\'\i guez
\cite{rod99a}; Verner et al.\ \cite{ver00}).
In any case, the [\ion{Fe}{ii}] lines measured in the near-infrared spectrum of
\object{M42} have been found to arise in regions of moderate density
(Marconi et al.\ \cite{mar98}; Luhman et al.\ \cite{luh98}), similar
to the densities measured from the usual diagnostics based on [\ion{S}{ii}],
[\ion{O}{ii}] or [\ion{Cl}{iii}] lines, and the [\ion{Fe}{ii}]~$\lambda$8617
intensity measured in \object{M42}, when compared with the intensities of
these infrared lines, implies a common origin at moderate densities (see, for
example, Fig.~5d in Bautista \& Pradhan \cite{bp98}).
Hence, the calculation of the $\element[+]{Fe}/\element[+]{H}$ abundance ratio
from [\ion{Fe}{ii}]~$\lambda$8617 for the physical conditions
$T_{\rm e}[\ion{N}{ii}]$ and $N_{\rm e}[\ion{S}{ii}]$ seems quite reasonable.

Another source of uncertainty in the values of $\element[+]{Fe}/\element[+]{H}$
in Table~2 arises from the atomic data used for \element[+]{Fe} -- particularly
the collision strengths.
The \element[+]{Fe} abundances have been calculated with the emissivities
derived by Bautista \& Pradhan (\cite{bp96}) using their own collision
strengths, although Bautista \& Pradhan (\cite{bp98}) consider more accurate
the collision strengths of Pradhan \& Zhang (\cite{pz93}) and Zhang \& Pradhan
(\cite{zp95}).
However, this other set of collision strengths would lead to very similar
values ($\sim$30\% lower) for the \element[+]{Fe} abundance derived from
[\ion{Fe}{ii}]~$\lambda$8617.
The collision strengths presented by Pradhan \& Zhang (\cite{pz93}) and Zhang
\& Pradhan (\cite{zp95}) can also be affected by uncertainties (Rodr\'\i guez
\cite{rod99a}; Oliva et al. \cite{omd99}), but the contribution of
\element[+]{Fe} to the total abundance is low and these uncertainties will not
affect significantly the derived \element{Fe} abundances.
The \element{Fe} abundance could be alternatively derived from the
values of $\element[++]{Fe}/\element[+]{H}$ and ionization correction factors.
This is the approach followed in Rodr\'\i guez (\cite{rod96}), and leads to
very similar results.

\section{\element{Fe} abundance}

\begin{figure}
  \resizebox{\hsize}{!}{\includegraphics{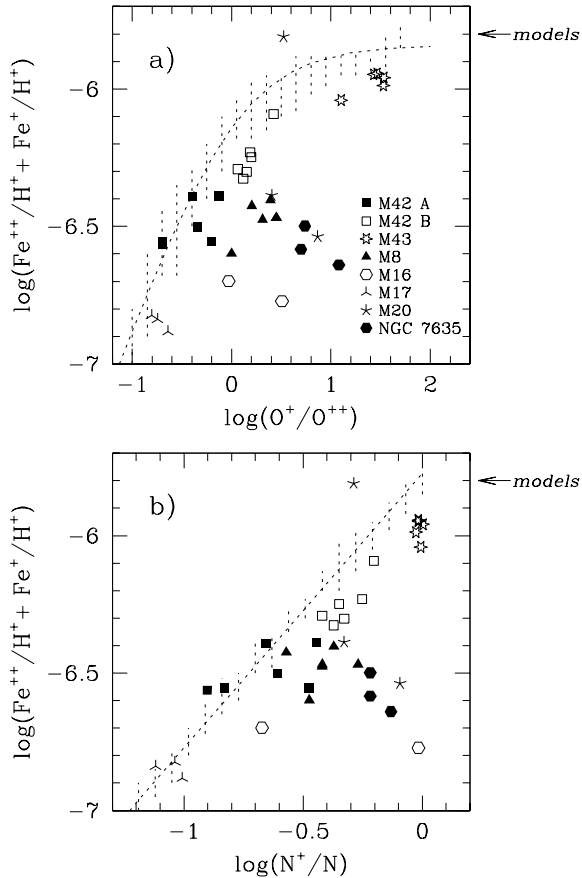}}
    \caption{The mean values of
        $\element[+]{Fe}/\element[+]{H}+\element[++]{Fe}/\element[+]{H}$
	are presented as  functions of the degree of ionization given by
	$\element[+]{O}/\element[++]{O}$ ({\bf a}) and
	$\element[+]{N}/\element{N}$ ({\bf b}), with
	$\log(\element{N}/\element{H})=-4.35$. The relations given by
	Eqs.(\ref{eq1}) and (\ref{eq2}) are also shown, scaled to
	$\log(\element{Fe}/\element{H})=-5.8$; The vertical dotted lines show
	the range covered by the individual models (references in the text).
	It has been assumed that for those positions where
	[\ion{Fe}{ii}]~$\lambda$8617 could not be measured (see Table~2),
	$\element[+]{Fe}/\element[+]{H}$ has a negligible contribution}
    \label{fig2}
\end{figure}

The values of $\log\left(\element[++]{Fe}/\element[+]{H}+
\element[+]{Fe}/\element[+]{H}\right)$ are presented as  functions of the
degree of ionization, $\log\left(\element[+]{O}/\element[++]{O}\right)$ and
$\log\left(\element[+]{N}/\element{N}\right)$, in Fig.~\ref{fig2}, where
$\log\left(\element{N}/\element{H}\right)=-4.35$ has been assumed to hold for
all the nebulae (Rodr\'\i guez \cite{rod99b}).
Figure~\ref{fig2} shows that
$\element[++]{Fe}/\element[+]{H}+\element[+]{Fe}/\element[+]{H}$ decreases with
increasing degree of ionization for the positions in \object{M42} and
\object{M17}, reflecting the expected increment in the contribution of
\element[3+]{Fe} to the total abundance.
However, the regions with lower degree of ionization -- excluding \object{M43} --
show unexpectedly low values of $\element[++]{Fe}/\element[+]{H}+
\element[+]{Fe}/\element[+]{H}$, that cannot be due to an error in the
derived $\element[+]{Fe}/\element[+]{H}$ abundance ratio, as argued above and
in Sect.~7.1.

Figures~\ref{fig2}a and \ref{fig2}b also show an indication of the expected --
i.e. model predicted -- behaviour of the ionization fractions
$x(\element[++]{Fe})+x(\element[+]{Fe})$ as a function of
$x(\element[+]{O})/x(\element[++]{O})$ and $x(\element[+]{N})$, scaled to
a total abundance $\log(\element{Fe}/\element{H})=-5.8$.
Two grids of ionization models have been used: those from Stasi\'nska
(\cite{sta90}) for solar metallicity $\mbox{Z}_\odot$ and for $\mbox{Z}_\odot/2$,
and those from Gruenwald \& Viegas (\cite{gru92}) for $\mbox{Z}_\odot$ and
$\mbox{Z}_\odot/3$ (see Rodr\'\i guez \cite{rod99b} for further details on the
selected models).
The vertical dotted lines in Figs.~\ref{fig2}a and \ref{fig2}b show the range
covered by the individual photoionization models; a further dotted line shows
the assumed ionization correction factors, also scaled in these figures to
$\log(\element{Fe}/\element{H})=-5.8$, and given by:

\begin{equation}
\element{Fe}/\element{H}=1.1\times\left(\element{O}/\element[+]{O}\right)
\times
\left(\element[++]{Fe}/\element[+]{H}+\element[+]{Fe}/\element[+]{H}\right),
\label{eq1}
\end{equation}

\begin{equation}
\element{Fe}/\element{H}=0.94\times\left(\element{N}/\element[+]{N}\right)
\times
\left(\element[++]{Fe}/\element[+]{H}+\element[+]{Fe}/\element[+]{H}\right).
\label{eq2}
\end{equation}

These relations provide the correction for the contribution of
\element[3+]{Fe} to the \element{Fe} abundance in terms of the ionic
and total abundances of \element{O} and \element{N}.
According to the models, Eqs.~(\ref{eq1}) and (\ref{eq2}) have accuracies
better than $\pm$0.2 and $\pm$0.1~dex, respectively.
\element[3+]{Fe}, \element[++]{O} and \element[++]{N} are formed at
30.6, 35.1 and 29.6~eV, and none of these ions is expected to be further
ionized.
Hence, the ionization correction factor based on \element{N} should be more
reliable, but whereas the total \element{O} abundance can be derived from the
\element[+]{O} and \element[++]{O} abundances using optical observations,
the total \element{N} abundance has to be estimated from \element[+]{N}.
Therefore, the results for both $\element{Fe}/\element{N}$ and
$\element{Fe}/\element{O}$ are presented here, and $\element{Fe}/\element{H}$
is obtained from $\element{Fe}/\element{O}\times\element{O}/\element{H}$.

The real accuracy of Eqs.(\ref{eq1}) and (\ref{eq2}) is difficult to estimate.
Even if the ionization and recombination cross-sections used by the models
are reliable, the predicted ionic concentrations may show systematic departures
from those prevailing in the observed objects due to the necessarily
unrealistic radiation fields and nebular structures assumed in the modeling.
The dependence of the results on the radiation field and the nebular structure
can be minimized by comparing, for observations and models, ions whose
ionization potentials are similar, as the ions of \element{O} and \element{N}
are compared with the \element{Fe} ions in Eqs.(\ref{eq1}) and (\ref{eq2}).
The dispersion shown by the models around these relations in Eqs.(\ref{eq1})
and (\ref{eq2}), may then reflect real second-order variations in the spectral
distribution of the radiation field, the gas density or its metallicity.
The concentrations of those ions with low ionization potentials (such as
\element[0]{O}, \element[+]{Fe} or even \element[+]{Cl} and \element[+]{He}) or
relatively high ones (such as
 \element[3+]{Ar}) are, however, severely affected by
the unrealistic radiation fields or nebular structures assumed by models, but
the relative ionization fractions of those ions whose concentrations depend on
the flux density
 of photons with energies of 30--35~eV -- as in Eqs.(\ref{eq1}) and
(\ref{eq2}) -- seem quite reliable (Rodr\'\i guez \cite{rod99b}).
Furthermore, the \element{Fe} abundances will be more reliable when obtained
from $\element[+]{Fe}/\element[+]{H}+\element[++]{Fe}/\element[+]{H}$ than from
$\element[++]{Fe}/\element[+]{H}$.
The relations implied by Eqs.(\ref{eq1}) and (\ref{eq2}) are also similar to
the results obtained from individual ionization models.
The photoionization models presented by Baldwin et al.\ (\cite{bald91}) and
Rubin et al.\ (\cite{rub91a}, \cite{rub91b}) for two regions in \object{M42}
agree with Eqs.(\ref{eq1}) and (\ref{eq2}) to within 0.25~dex.
Finally, the model derived by Bautista \& Pradhan (\cite{bp98}) also for
\object{M42}, which includes the most recent values for the photoionization
cross-sections and recombination-rate coefficients for the \element{Fe} ions,
agrees more closely with Eq.(\ref{eq1}).
In any case, to confirm the trends in Eqs.(\ref{eq1}) and (\ref{eq2}),
further calculations involving these new ionization and recombination
cross-sections for different model assumptions would be of interest.

\begin{figure}
  \resizebox{\hsize}{!}{\includegraphics{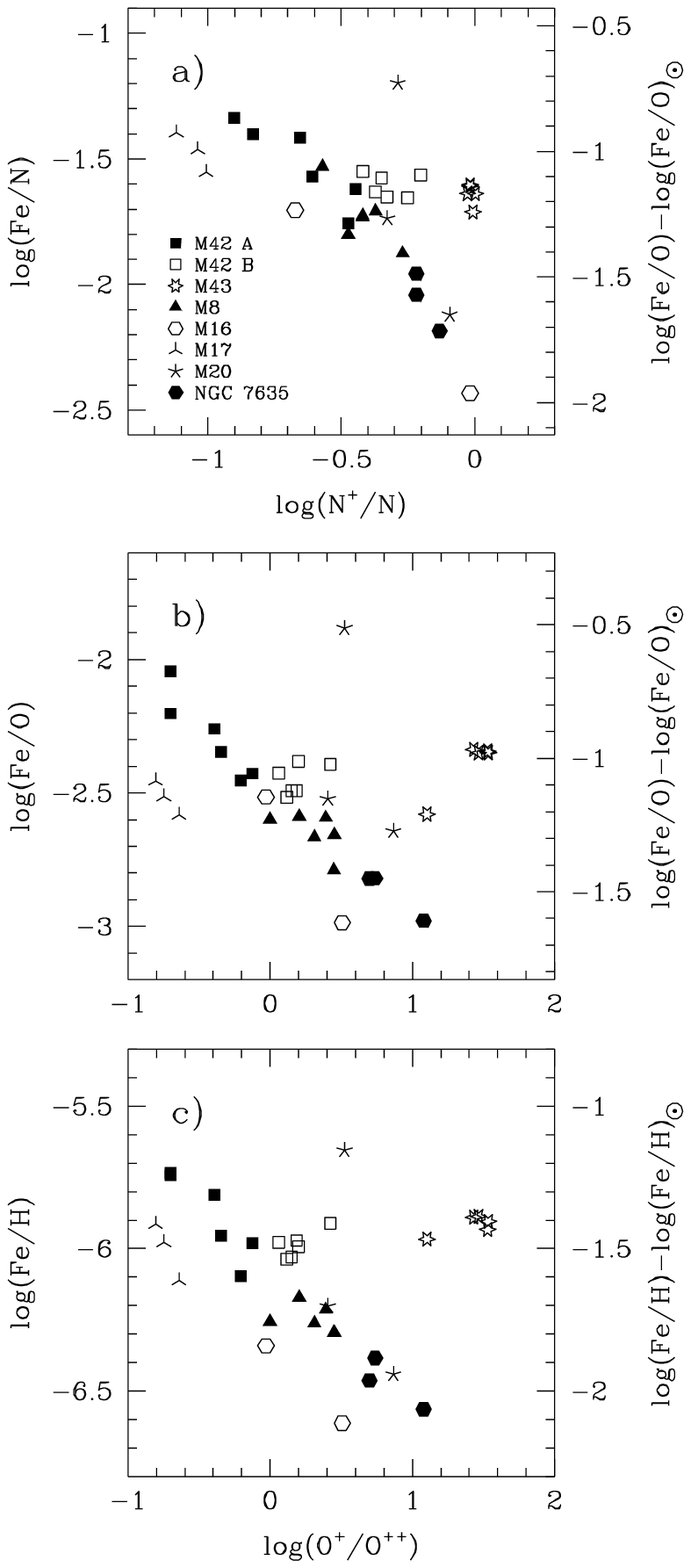}}
    \caption{The values of $\element{Fe}/\element{N}$ ({\bf a}),
	$\element{Fe}/\element{O}$ ({\bf b}) and $\element{Fe}/\element{H}$
	({\bf c}), as derived from Eq.(\ref{eq1}), Eq.(\ref{eq2}) and the
	relation $\element{Fe}/\element{H}=\element{Fe}/\element{O}\times
	\element{O}/\element{H}$. The values of $\element{Fe}/\element{H}$
	implied by the $\element{Fe}/\element{N}$ abundance ratio can be
	obtained from: $\log(\element{Fe}/\element{H})=
	\log(\element{Fe}/\element{N})-4.35$ (Rodr\'\i guez \cite{rod99b})}
    \label{fig3}
\end{figure}

The values derived for $\element{Fe}/\element{N}$, $\element{Fe}/\element{O}$
and $\element{Fe}/\element{H}$ are listed in Table~3 and shown as a function
of degree of ionization in Fig.~\ref{fig3}.
The results are compared with the solar abundances in the right hand axes of
Fig.~\ref{fig3}, where the solar values are
$\log(\element{Fe}/\element{N})_\odot=-0.47$,
$\log(\element{Fe}/\element{O})_\odot=-1.37$ and
$\log(\element{Fe}/\element{H})_\odot=-4.50$ (Grevesse et al.\ \cite{gre96}).
The depletion factors implied by the $\element{Fe}/\element{N}$ and
$\element{Fe}/\element{O}$ abundance ratios are broadly consistent, especially
if it is taken into account that whereas no significant amount of \element{N}
is expected to be present in dust (Sofia et al.\ \cite{sof94}), about 25\% of
the \element{O} abundance ($\sim0.1$~dex) may be depleted in refractory dust
grains (Cardelli et al.\ \cite{car96}).
However, the depletion factors implied by $\element{Fe}/\element{H}$ are much
lower, probably reflecting the difficulties related to the choice of the
correct reference abundances.
Several recent studies of the abundances of different elements in the
interstellar medium and in B~stars suggest that the abundances of \element{O},
\element{C}, \element{N} and \element{Kr} relative to \element{H} are $\sim2/3$
the solar values (e.g. Snow \& Witt \cite{sno96}; Cardelli \& Meyer
\cite{car97}; Sofia et al.\ \cite{sof97}), and similar underabundances for
\element{O}, \element{N} and other elements are found in Rodr\'\i guez
(\cite{rod99b}) for all the objects studied here.
If this result holds also for the $\element{Fe}/\element{H}$ abundance ratio,
the most reliable depletion factors must be those implied by the
$\element{Fe}/\element{N}$ or $\element{Fe}/\element{O}$ abundance ratios.

The \element{Fe} abundances presented here are in general about 0.2~dex lower
than the values derived by other authors (e.g. Osterbrock et al.\ \cite{otv92}
for \object{M42}; Esteban et al.\ \cite{est98}, \cite{est99a} for \object{M42}
and \object{M8}; Rodr\'\i guez \cite{rod96}), reflecting the different atomic
data and the slightly different ionization correction factors used.
As an example, if the line intensities measured by Osterbrock et al.\
(\cite{otv92}) in \object{M42} are analysed with the same atomic data used
here and the same procedure followed, their derived
$\log(\element{Fe}/\element{H})=-5.57$ (for
$\element[+]{O}/\element[++]{O}\sim1$) would change to
$\log(\element{Fe}/\element{H})=-5.83$ if their ionization correction factor,
$\element[3+]{Fe}/\element[++]{Fe}=2$, is used, and to
$\log(\element{Fe}/\element{H})=-5.92$ if Eq.(\ref{eq1}) is used instead.

The values of the \element{Fe} abundance listed in Table~3 are highly sensitive
to the [\ion{Fe}{iii}] atomic data (in particular the collision strengths)
and to the assumed ionization correction factors.
Zhang (\cite{zha96}) estimates uncertainties of 10--20\% for the collision
strengths used here to derive the \element[++]{Fe} abundances so that, even if
the uncertainties are somewhat higher, the contribution of \element[3+]{Fe} to
the total abundance seems to be the main source of uncertainty.

New atomic data for the ionization and recombination rates of the \element{Fe}
ions, or for the rates of their charge-exchange reactions, could change
the predicted variation of $x(\element[+]{Fe})+x(\element[++]{Fe})$ with
$x(\element[+]{O})/x(\element[++]{O})$ or $x(\element[+]{N})$, but the effects
can only be significant for the higher ionization regions where the
contribution of \element[3+]{Fe} to the total abundance dominates.
For those regions, namely the positions in \object{M42}~A and \object{M17},
it might not be surprising to find \element{Fe} abundances similar to those in
\object{M42}~B, so that the trend in Fig.~\ref{fig3} would level off for
$\element[+]{O}/\element[++]{O}<1$.
However, the trend followed by the \element{Fe} abundances in the regions of
lower ionization is established by the values derived for
$\element[++]{Fe}/\element[+]{H}+\element[+]{Fe}/\element[+]{H}$ (see
Fig.~\ref{fig2}) and seems quite robust.
As discussed above, the real values of the depletion factors implied by the
derived abundances, are somewhat uncertain, but nevertheless they imply that
most of the \element{Fe} atoms remain in the dust.
The different degrees of depletion factors found for the different objects seem
to be well established, and the dependence of the results on the ionization
degree offers an explanation, since it relates those photons with energies
$\ga35$~eV to the release of \element{Fe} atoms from dust\footnote{The
degree of ionization is proportional to the number of energetic photons
per electron (Osterbrock \cite{ost89}), and since the distributions of gas
and dust within a nebula are expected to follow a roughly similar pattern
(see M\"unch and Persson \cite{mun71} for \object{M42}), a higher gas
density is likely to be associated with a higher dust concentration.
Thus, the correlation can be interpreted as a relationship between the number of
energetic photons available per dust grain and the amount of \element{Fe}
released from dust.}.
The implications of this result will be discussed in Sect.~8.

\subsection{Comments on \object{M43} and \object{M20}--1}

The \element{Fe} abundance ratios derived for \object{M43} and \object{M20}--1
show significant deviations from the relationship with the degree of ionization
followed by the other objects.
\object{M43} is the only region in the sample which is excited by a B-type star
(\object{NU~Ori}), its spectrum is greatly affected by dust-scattered stellar
light, and is located several arc minutes to the north of \object{M42},
belonging to the same molecular complex, although it is an independent
\ion{H}{ii} region (Goudis \cite{gou82}).
The $\element{Fe}/\element{H}$ abundance ratio in \object{M43} seems reliable
since it is given by the calculated values for
$\element[+]{Fe}/\element[+]{H}$ and $\element[++]{Fe}/\element[+]{H}$ and does
not depend on ionization correction factors.
A contribution of fluorescence effects, which are very important in
\object{M43} (Rodr\'\i guez \cite{rod99a}), to the [\ion{Fe}{ii}]~$\lambda8617$
intensity cannot be an explanation, since when
the \element{Fe} abundances are derived just from [\ion{Fe}{iii}] emission,
which is not expected to suffer from fluorescence effects (Lucy \cite{lucy95};
Bautista \& Pradhan \cite{bp98}),
the same deviation is found for \object{M43} (Rodr\'\i guez \cite{rod96}).
The relatively high \element{Fe} abundance in this object
is not readily explained; maybe it is just an indication of the dependence of
the \element{Fe} depletion on other factors apart from the radiation field:
age, density, previous history, etc.

\begin{table}
\caption[ ]{\element{Fe} abundances}
\begin{tabular}{llll}
\hline
\noalign{\smallskip}
Object & $\element{Fe}/\element{N}$ & $\element{Fe}/\element{O}$ &
	$\element{Fe}/\element{H}^{\rm a}$ \\
\noalign{\smallskip}
\hline
\noalign{\smallskip}
\object{M42} A--1   &2.4$\times10^{-2}$&3.7$\times10^{-3}$&1.0$\times10^{-6}$\\
\object{M42} A--2   &1.7$\times10^{-2}$&3.5$\times10^{-3}$&8.0$\times10^{-7}$\\
\object{M42} A--3   &4.6$\times10^{-2}$&6.3$\times10^{-3}$&1.8$\times10^{-6}$\\
\object{M42} A--4   &2.7$\times10^{-2}$&4.5$\times10^{-3}$&1.1$\times10^{-6}$\\
\object{M42} A--5   &3.8$\times10^{-2}$&5.5$\times10^{-3}$&1.5$\times10^{-6}$\\
\object{M42} A--6   &4.0$\times10^{-2}$&9.0$\times10^{-3}$&1.8$\times10^{-6}$\\
\noalign{\smallskip}
\hline
\noalign{\smallskip}
\object{M42} B--1   &2.2$\times10^{-2}$&3.2$\times10^{-3}$&9.4$\times10^{-7}$\\
\object{M42} B--2   &2.8$\times10^{-2}$&3.7$\times10^{-3}$&1.0$\times10^{-6}$\\
\object{M42} B--3   &2.3$\times10^{-2}$&3.1$\times10^{-3}$&9.2$\times10^{-7}$\\
\object{M42} B--4   &2.7$\times10^{-2}$&4.1$\times10^{-3}$&1.2$\times10^{-6}$\\
\object{M42} B--5   &2.7$\times10^{-2}$&4.1$\times10^{-3}$&1.0$\times10^{-6}$\\
\object{M42} B--6   &2.4$\times10^{-2}$&3.2$\times10^{-3}$&1.1$\times10^{-6}$\\
\noalign{\smallskip}
\hline
\noalign{\smallskip}
\object{M43}--1     &2.3$\times10^{-2}$&4.5$\times10^{-3}$&1.2$\times10^{-6}$\\
\object{M43}--2     &2.6$\times10^{-2}$&4.6$\times10^{-3}$&1.3$\times10^{-6}$\\
\object{M43}--3     &2.1$\times10^{-2}$&2.6$\times10^{-3}$&1.1$\times10^{-6}$\\
\object{M43}--4     &2.3$\times10^{-2}$&4.2$\times10^{-3}$&1.2$\times10^{-6}$\\
\object{M43}--5     &2.5$\times10^{-2}$&4.5$\times10^{-3}$&1.3$\times10^{-6}$\\
\noalign{\smallskip}
\hline
\noalign{\smallskip}
\object{M8}--1      &1.9$\times10^{-2}$&2.2$\times10^{-3}$&5.1$\times10^{-7}$\\
\object{M8}--2      &1.3$\times10^{-2}$&1.6$\times10^{-3}$&5.1$\times10^{-7}$\\
\object{M8}--3      &2.0$\times10^{-2}$&2.6$\times10^{-3}$&6.1$\times10^{-7}$\\
\object{M8}--4      &2.0$\times10^{-2}$&2.2$\times10^{-3}$&5.5$\times10^{-7}$\\
\object{M8}--5      &2.9$\times10^{-2}$&2.6$\times10^{-3}$&6.7$\times10^{-7}$\\
\object{M8}--6      &1.6$\times10^{-2}$&2.5$\times10^{-3}$&5.5$\times10^{-7}$\\
\noalign{\smallskip}
\hline
\noalign{\smallskip}
\object{M16}--1     &3.7$\times10^{-3}$&1.0$\times10^{-3}$&2.4$\times10^{-7}$\\
\object{M16}--2     &2.0$\times10^{-2}$&3.1$\times10^{-3}$&4.6$\times10^{-7}$\\
\noalign{\smallskip}
\hline
\noalign{\smallskip}
\object{M17}--1     &2.8$\times10^{-2}$&2.6$\times10^{-3}$&7.7$\times10^{-7}$\\
\object{M17}--2     &3.4$\times10^{-2}$&3.5$\times10^{-3}$&1.2$\times10^{-6}$\\
\object{M17}--3     &4.0$\times10^{-2}$&3.1$\times10^{-3}$&1.0$\times10^{-6}$\\
\noalign{\smallskip}
\hline
\noalign{\smallskip}
\object{M20}--1     &6.3$\times10^{-2}$&1.3$\times10^{-2}$&2.2$\times10^{-6}$\\
\object{M20}--2     &1.8$\times10^{-2}$&3.0$\times10^{-3}$&6.3$\times10^{-7}$\\
\object{M20}--3     &7.6$\times10^{-3}$&2.3$\times10^{-3}$&3.6$\times10^{-7}$\\
\noalign{\smallskip}
\hline
\noalign{\smallskip}
\object{NGC~7635}--1&6.5$\times10^{-3}$&1.0$\times10^{-3}$&2.7$\times10^{-7}$\\
\object{NGC~7635}--2&1.1$\times10^{-2}$&1.5$\times10^{-3}$&4.1$\times10^{-7}$\\
\object{NGC~7635}--3&9.1$\times10^{-3}$&1.5$\times10^{-3}$&3.4$\times10^{-7}$\\
\noalign{\smallskip}
\hline
\noalign{\smallskip}
\multicolumn{4}{l}{$^{\rm a}$ Derived from $\element{Fe}/\element{O}\times
	\element{O}/\element{H}$}\\
\end{tabular}
\end{table}

The deviation of \object{M20}--1 from the relationships in Fig.~\ref{fig3}
could be precisely reflecting the action of another factor.
Position \object{M20}--1 is located at the tip of \object{HH~399}, as mentioned
above, a jet extending into the ionized gas from a molecular column in
\object{M20} (Cernicharo et al.\ \cite{cer98}; Hester et al.\ \cite{hes99}).
The jet is fully ionized and its spectrum is very similar to those measured for
\object{M20}--2 and \object{M20}--3, yielding similar values for the physical
conditions, chemical abundances and ionization fractions for all elements
excluding \element{Fe} (Rodr\'\i guez \cite{rod99b}).
In fact, the only features showing distinct behaviour in the spectrum of
\object{M20}--1 are the [\ion{Fe}{iii}], [\ion{Fe}{ii}] and [\ion{Ni}{ii}]
lines, whose intensities relative to $\element{H}\beta$ are significantly
higher than those measured for \object{M20}--2 and \object{M20}--3
(see Tables~1 and 2).
Even the relative intensities of the [\ion{Fe}{ii}] lines, whose values reflect
the effects of fluorescence in most [\ion{Fe}{ii}] lines (Rodr\'\i guez
\cite{rod96}, \cite{rod99a}), depart in \object{M20}--1 from the line ratios
measured for the other objects.
For example, the intensity ratio between [\ion{Fe}{ii}]~$\lambda$4287, expected
to be very sensitive to fluorescence, and [\ion{Fe}{ii}]~$\lambda$8617, almost
insensitive to pumping effects, is found to be lower in \object{M20}--1 than in
any other position, suggesting that fluorescence effects are minimal in this
position.
Therefore, although [\ion{Fe}{ii}]~$\lambda$8617 could in principle be affected
by fluorescence in some objects, the high values derived for
$\element[+]{Fe}/\element[+]{H}$ and $\element[+]{Fe}/\element[++]{Fe}$ in
\object{M20}--1 (the value of this latter ratio being above those found for
lower-excitation objects like \object{M43} and \object{NGC~7635}),
cannot be explained by resonance effects.
The [\ion{Fe}{ii}]~$\lambda$8617 line could only be measured in another
position in this nebula (\object{M20}--2), where
$\element[+]{Fe}/\element[++]{Fe}$ is found to be equally high, but since this
line intensity is highly uncertain in \object{M20}--2, it is not clear whether
these values of $\element[+]{Fe}/\element[++]{Fe}$ are characteristic of
\object{M20} or are peculiar to the jet sampled in \object{M20}--1 and imply
a higher dust destruction efficiency in the lower ionization layers of the jet.
In any case, a similar, although less extreme, \element{Fe} overabundance in
\object{M20}--1 can be found when the \element{Fe} abundances are calculated
from just [\ion{Fe}{iii}] emission and model-predicted ionization fractions
(Rodr\'\i guez \cite{rod96}).
This overabundance of \element{Fe} atoms in the gas, which has also been found
for several typical, not fully ionized HH objects
(Beck-Winchatz et al.\ \cite{bec96} and references therein), presumably
reflects dust destruction due to shock waves arising
from the interaction with the ionized gas of the material in the jet, which is
estimated to move at a velocity of $\sim370~\mbox{km~s}^{-1}$ (Hester et al.\
\cite{hes99})\footnote{See also http://eagle.la.asu.edu/hester/m20.html}.

\subsection{{\rm [\ion{Fe}{iv}]}~$\lambda2837$ and \element[3+]{Fe} abundance}

The only available measurement of an [\ion{Fe}{iv}] line in an \ion{H}{ii}
region is that performed by Rubin et al.\ (\cite{rub97}) for
[\ion{Fe}{iv}]~$\lambda2837$ in \object{M42}.
Their measured reddening-corrected intensity ratio,
$I(2837)/I(\element{H}\beta)\simeq10^{-3}$, along with two previous
photoionization models for two regions in \object{M42} and the collision
strengths calculated by Berrington \& Pelan (\cite{ber95}, \cite{ber96}) for
\element[3+]{Fe}, imply $\element{Fe}/\element{H}=4.6\times10^{-7}$ and
$1.6\times10^{-7}$ (with the difference between the two models' results being
due mostly to their different predicted temperatures).
These abundance ratios are significantly below
$\element{Fe}/\element{H}=3.0\times10^{-6}$, the value Rubin et al.\
(\cite{rub97}) consider consistent with [\ion{Fe}{ii}] and [\ion{Fe}{iii}]
emission in \object{M42} and their models for this nebula.

The discrepancy is somewhat reduced if the abundances derived here from
[\ion{Fe}{ii}] and [\ion{Fe}{iii}] lines (and more recent atomic data for
[\ion{Fe}{iii}]) in \object{M42} are considered:
$\element{Fe}/\element{H}=(0.8\mbox{--}1.8)\times10^{-6}$.
Although the weakness of the only measured [\ion{Fe}{iv}] line, its sensitivity
to temperature and the extinction correction may account for part of the
discrepancy, it could be entirely due to errors in the collision strengths for
\element[3+]{Fe}.
However, the more extensive calculations of the collision strengths for this
ion performed by Zhang \& Pradhan (\cite{zp97}) lead to very similar
abundances.
A 33-level model atom of \element[3+]{Fe} based on the collision strengths of
Zhang \& Pradhan (\cite{zp97}) and the transition probabilities from Garstang
(\cite{gar58}), for temperatures in the range 8000--8500~K and densities of
several thousands (these physical conditions being in agreement with those
usually measured in different zones in \object{M42} from [\ion{O}{iii}] and
[\ion{Cl}{iii}] lines that should originate in the [\ion{Fe}{iv}] emitting
region: e.g. McCall \cite{mcc79}; Torres-Peimbert et al.\ \cite{tor80};
Esteban et al. \cite{est98}; Rodr\'\i guez \cite{rod99b}) and the intensity
ratio measured for [\ion{Fe}{iv}]~$\lambda2837$ by Rubin et al.\
(\cite{rub97}), gives $\element[3+]{Fe}/\element[+]{H}\simeq2\times10^{-7}$.
This value is a factor of 2--8 lower than the \element[3+]{Fe} abundances
implied by the results presented here for \object{M42} in Fig.~\ref{fig3}c and
Table~3: $\element[3+]{Fe}/\element[+]{H}=(4\mbox{--}15)\times10^{-7}$.
The discrepancy is merely of a factor of $\sim2$ if only \object{M42}~A--1 and
\object{M42}~A--2, located only 8\farcs5 south of the position observed by
Rubin et al.\ (\cite{rub97}), are considered.
Given all the uncertainties related to the determination of
$\element[3+]{Fe}/\element[+]{H}$ from just one UV, weak [\ion{Fe}{iv}] line
measured in only one position whose optical spectra and [\ion{Fe}{ii}] and
[\ion{Fe}{iii}] line intensities are not available at the moment, and given the
wide range covered by the values derived here for $\element{Fe}/\element{H}$
in \object{M42} and their uncertainties, a discrepancy of a factor $\ge2$ might
not be significant.

Taken at face value, the discrepancy could indicate a failure in the model
predictions for the \element{Fe} ionization fractions and a corresponding
uncertainty in the \element{Fe} abundances derived for high-ionization
\ion{H}{ii} regions.
It has also been suggested that the lower \element{Fe} abundance implied by the
[\ion{Fe}{iv}] line could be real, reflecting the presence of a gradient in the
gaseous abundance of this element (Bautista \& Pradhan \cite{bp98}), but the
uncertainties related to the determination of the \element[3+]{Fe} abundance
are currently too high to reach any definite conclusion.
Further studies of [\ion{Fe}{iii}] and [\ion{Fe}{ii}] lines at the position
where [\ion{Fe}{iv}]~$\lambda2837$ has been observed and an assessment of the
accuracy of the atomic data for \element[3+]{Fe} will be needed to reach a full
understanding of [\ion{Fe}{iv}] emission in \ion{H}{ii} regions.

\section{\element{Fe} abundance, extinction and dust in \ion{H}{ii} regions}

\subsection{\element{Fe} abundance and dust destruction}

The high depletion factors measured for \element{Fe} in the interstellar
medium and the high cosmic abundance of this element imply that \element{Fe}
must contribute a substantial amount of mass to the dust grains (Sofia et al.\
\cite{sof94}).
Therefore, the degree of incorporation of \element{Fe} atoms to dust should
give a good indication of the dust-to-gas ratio.
Since the seven \ion{H}{ii} regions studied here have similar chemical
abundances for those elements usually found in the gas phase (Rodr\'\i guez
\cite{rod99b}), it will be assumed that the \element{Fe} total abundance, i.e.
in gas and dust, is the same for all these objects.
The \element{Fe} abundance values derived here can then be compared to a
reference \element{Fe} abundance to determine the fraction of \element{Fe}
atoms present in the gas and depleted into dust for each area considered.
Taking the solar \element{Fe} abundance as reference, the values derived for
$\element{Fe}/\element{H}$ or $\element{Fe}/\element{N}$ imply that about
0.8--6\% or 1--19\% of the \element{Fe} atoms have been released to the gas
phase.
The corresponding logarithmic depletion factors, shown in Fig.~\ref{fig3},
fall between the values typically found for cool clouds in the Galactic disc,
where $\log(\element{Fe}/\element{H})-\log(\element{Fe}/\element{H})_\odot
\simeq-2.09$ to $-2.27$, and those measured in the Galactic halo, which
are in the range $\simeq-0.58$ to $-0.69$ (Savage \& Sembach \cite{sav96}).
The relatively high \element{Fe} abundances in the halo are similar to the
values found for high-velocity clouds, and this difference in the \element{Fe}
depletion factors for cool disc and halo
clouds (and the intermediate values found for warm disc clouds and clouds
sampling both the disc and the halo) is interpreted as reflecting the
action of the primary dust destruction agent: shock waves due to supernovae.

The \element{Fe} abundances derived here imply that most of \element{Fe}
remains in the dust grains inside \ion{H}{ii} regions, but also imply the
release to the gas of a significant amount of these atoms from the dust grains
originally located inside molecular clouds, where up to 100\% of \element{Fe}
is expected to be in the solid phase.
Energetic photons, instead of shock waves, (excluding \object{M20}--1, see
Sect.~7.1) seem to be responsible for this dust destruction effect, as
suggested by the correlation of the \element{Fe} abundances with the
degree of ionization in Fig.~\ref{fig3}.
Two processes might be responsible for this effect: vaporization and optical
erosion or photodesorption, both leading to the release of atoms or molecules
from the grains after the absorption of photons.
Optical erosion results from the ionization or excitation of a bonding electron
of a molecule after the absorption of a photon; this mechanism is expected to
destroy only the ice mantles on the grains, and the refractory grains should
not be affected (Osterbrock \cite{ost89}).
[However, note that photodesorption cross-sections for metal atoms on dust
grains are not known, and this process could be of importance.]
Assuming then that sublimation is the process that releases \element{Fe} atoms
to the gas, the grains containing these atoms cannot be classical, i.e.
characterized by a homogeneous composition and structure and with radii
$a\sim0.1~\mu\mbox{m}$, since these classical particles are in thermal
equilibrium with the radiation field and have temperatures $T\sim100$~K
according to the calculations and the estimates obtained from the observed
infrared emission, whereas refractory particles have much higher evaporation
temperatures: $T\ge1000$~K.
However, if the grains are small enough, $a\le0.03~\mu\mbox{m}$, they can
undergo significant temperature fluctuations after the absorption of one or a
few energetic photons.
Several studies on the temperature fluctuations undergone by dust grains in
different environments conclude that only very small grains, with
$a\sim10~\mbox{\AA}$, reach temperatures high enough to be destroyed.
As an example, Guhathakurta \& Draine (\cite{guh89}) estimate that graphite or
silicate grains with $a\sim5~\mbox{\AA}$ located at 0.3~pc from a B3V star
could reach temperatures $T\sim2000$~K, would lose atoms through
sublimation and have lifetimes of about $3\times10^5$~years (whereas the ages
of \ion{H}{ii} regions are of the order of one million years).
These small dust grains contain about one thousandth of the total dust
mass (Sellgren \cite{sel84}) and their destruction cannot explain the
relatively high \element{Fe} abundances measured for \object{M42}.
Nevertheless, a further population of small grains could be located in larger
composite grains.
Interstellar grains, and in particular those from molecular clouds, are
expected to consist of loose aggregates of smaller subunits formed by an
accretion process that might fragment once inside the ionized gas or at the
ionization front.

As an example, PAH-like clusters within larger amorphous carbon grains could
be easily released.
Note that PAHs -- polycyclic aromatic hydrocarbons, molecules of size
$\sim10$~\AA\ proposed to explain several emission bands observed in different
environments such as the ionization fronts in \ion{H}{ii} regions -- are destroyed
inside \ion{H}{ii} regions (Bregman et al.\ \cite{bre89}; Giard et al.\
\cite{gia94}).
PAHs or other kinds of aromatic molecules could aggregate and form
conglomerates or amorphous carbon grains inside
dense molecular clouds, and the abundant and chemically active \element{Fe}
atoms might react with PAHs and accelerate the process.
In fact, organometallic complexes could contain 5 to 10\% of metal atoms in the
interstellar medium and, thus, these complexes would play an important role in
the depletion of heavy elements and in dust formation processes in the
interstellar medium (Serra et al.\ \cite{ser92}; Klotz et al.\ \cite{klo95}).

\subsection{\element{Fe} abundance and colour excess}

\begin{figure}
  \resizebox{\hsize}{!}{\includegraphics{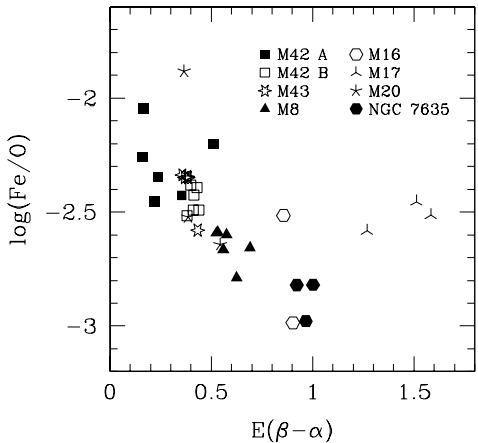}}
    \caption{The values of $\element{Fe}/\element{O}$ as a function of the
    colour excess derived from the relative intensities of $\element{H}\alpha$
    and $\element{H}\beta$}
    \label{fig4}
\end{figure}

The $\element{Fe}/\element{O}$ abundance ratios are shown as a
function of the colour excess implied by the relative intensities of the
\ion{H}{i} lines in Fig.~\ref{fig4}, where it can be seen that there is
a loose correlation between the two quantities.
The correlation of $\element{Fe}/\element{H}$ and $\element{Fe}/\element{N}$
with $E(\beta-\alpha)$ is also apparent, although somewhat less markedly than
that of $\element{Fe}/\element{O}$.
Previously, Rodr\'\i guez (\cite{rod96}) presented the correlation of
$\element{Fe}/\element{O}$ with $\tau_\beta$ (the effective optical depth
for extinction at $\element{H}\beta$), but the colour excess, which is an
observed quantity, independent of the extinction law, is a more
reliable and straightforward variable.


The positions in \object{M17} do not follow the correlation in
Fig.~\ref{fig4}, but this region is known to be heavily reddened by
foreground  dust, both interstellar and associated with the complex:
Chini \& Wargau (\cite{chi98}) estimate that $E(B-V)\sim1$~mag is due to the
diffuse interstellar medium in the line of sight to \object{M17} (note that
$E(\beta-\alpha)\simeq E(B-V)$ when $R_{V}=A_{V}/E(B-V)\simeq3$, the
typical value for the diffuse interstellar medium) whereas
the internal stars are further affected by $A_{V}>3$~mag of local
extinction.
The positions studied in the other \ion{H}{ii} regions might be only slightly
affected by external extinction due to a selection effect, since the
observations were centred on the brightest positions of objects with high
superficial brightness, but the difficulties inherent in disentangling the
extinction components due to external and internal dust, preclude an
estimate of the effects of interstellar extinction (i.e. unrelated to the
complex) in the correlation presented in Fig.~\ref{fig4}.

The extinction due to the interstellar medium along the line of sight to an
\ion{H}{ii} region can be estimated by studying the stars projected on the
nebula and assuming that those whose extinction law is similar to that usually
found for the diffuse interstellar medium should be located in front of the
stellar formation region.
This is the method followed by Chini \& Wargau (\cite{chi98}) for \object{M17},
and it has also been applied to \object{M42}, where Breger et al.\
(\cite{bre81}) find that $E(B-V)\simeq0.05$~mag is due to interstellar
extinction.
The distinction between extinction due to internal dust and that due to
external, associated dust is more difficult to make.
However, taking into account the effects of scattering by dust internal to
\ion{H}{ii} regions, this internal dust mixed with the ionized gas can produce
a maximum colour excess $E(\beta-\alpha)\simeq0.19$~mag (G\'omez Garrido \&
M\"unch \cite{gom84}).
Therefore, the correlation in Fig.~\ref{fig4} relates the \element{Fe}
abundance in the ionized gas -- arising from dust destruction taking place in
the region considered  -- with a colour excess necessarily due to foreground
dust.

Given a volume of gas and dust, the spatial separation between the two
components could be brought about through the effects of radiation pressure or
stellar winds on the dust grains.
However, collisional interactions and Coulomb forces between ions and charged
particles tend to hold dust and gas together in \ion{H}{ii} regions.
The efficiency of these processes depends on the size, structure and
composition of the grains; the available calculations show that classical
particles are strongly coupled to the gas (Osterbrock \cite{ost89}).
Small grains could be an exception: Cardelli \& Clayton (\cite{car88}) estimate
that grains with sizes $a=0.01~\mu\mbox{m}$ initially located at a distance $r$
from the central star can move under the effects of radiation pressure by
$\Delta r/r\ge0.2$ in $10^4$~years.
Since many \ion{H}{ii} regions are known to have complex structures, with dense
clumps or molecular columns inside or in front of the ionized gas (e.g.
\object{M8} and \object{NGC~7635}, as shown by their detailed HST images),
the overall effect of radiation pressure
or stellar winds could well be to push a significant amount of dust outside the
nebula into our line of sight.

In conclusion, both correlations shown by the gaseous \element{Fe} abundance
in \ion{H}{ii} regions seem to require the presence of dust grains differing
from the classical grains due to their smaller sizes or loose structure.
These non-classical grains could contain 10--20\% of the total \element{Fe}
abundance, their destruction would be due to photons with energies above 35~eV
and could take place after the fragmentation of an original composite grain.
On the other hand, on a longer timescale, the surviving non-classical grains
would be pushed outside the ionized region through the action of radiation
pressure or stellar winds, where they would be responsible for most of the
colour excess found in the objects studied here.

\section{Conclusions}

The spectra of several positions inside seven Galactic \ion{H}{ii} regions
have been analysed to derive their \element{Fe} abundance.
The relative intensities measured for the [\ion{Fe}{iii}] lines have been
compared with the values predicted by three combinations of the available
atomic data, and it has been shown that the collision strengths of Zhang
(\cite{zha96}) and the transition probabilities of Quinet (\cite{qui96}) lead
to the best agreement between the observed and predicted line ratios.
The $\element[++]{Fe}/\element[+]{H}$ abundance ratios derived for all the
positions studied with the above mentioned atomic data are 10--50\% lower than
the values implied by the calculations of Keenan et al.\ (\cite{kee92}), which
are based on older atomic data and have been used in several previous studies
of the \element{Fe} abundance in nebulae.

The $\element[+]{Fe}/\element[+]{H}$ abundance ratios have been calculated
using the [\ion{Fe}{ii}]~$\lambda8617$ line, which is almost insensitive to
fluorescence, and the emissivities of Bautista \& Pradhan (\cite{bp96}).
The derived \element[+]{Fe} abundances may be affected by uncertainties in the
measured intensities and in the collision strengths, but the effects of these
uncertainties in the \element{Fe} abundances should be small, since the
contribution of \element[+]{Fe} to the total abundance is very low for most
positions.
The \element{Fe} abundances are consequently derived from the values of
$\element[++]{Fe}/\element[+]{H}+\element[+]{Fe}/\element[+]{H}$ and ionization
correction factors for the contribution of \element[3+]{Fe} obtained from
available grids of photoionization models.

The derived $\element{Fe}/\element{O}$ abundance ratios are 3 to 40 times lower
than the solar value, implying that most of the \element{Fe} atoms remain
in dust grains inside the ionized volume.
These underabundance factors are, however, significantly lower than those found
for the undisturbed interstellar medium, where more than 99\% of \element{Fe}
is depleted into dust (Savage \& Sembach \cite{sav96}).
The correlation between the calculated \element{Fe} abundances and the
degree of ionization explains this difference by relating the energetic photons in
\ion{H}{ii} regions with the release of \element{Fe} atoms from dust.
Further agents can be responsible for dust destruction in \ion{H}{ii} regions,
and a remarkable example is provided by position \object{M20}--1, on the tip of
the optical jet \object{HH~399}, where the \element{Fe} gaseous abundance is
significantly higher than in the other areas, presumably indicating dust
destruction due to the shock waves related to the jet.

The inferred dust destruction by photons, together with the correlation of the
\element{Fe} abundance with the colour excess, can be explained by invoking
the presence of loosely-bound dust grains or mantles, containing 10 to 20\% of
the \element{Fe} abundance, whose fragmentation would produce grains small
enough to be destroyed by sublimation processes, whereas
the remnants of these small grains would be swept out of the ionized region by
radiation pressure or stellar winds.

The present results rely on the estimated contribution of \element[3+]{Fe}
to the total \element{Fe} abundance. The uncertainties related to this
estimate, strengthened by the current difficulties in explaining
[\ion{Fe}{iv}] emission in M42, may cast some doubts on the validity of the
results for the high ionization objects.
Certainly, this issue should be resolved before the correlation between the
\element{Fe} abundances and the degree of ionization can be considered as
firmly established, thus favouring the proposed interpretation.
If such were not the case, and the predicted contribution of \element[3+]{Fe}
to the total abundance were found to be grossly in error, the new results would
surely lead to alternative scenarios for the evolution of dust in \ion{H}{ii}
regions.

\begin{acknowledgements}
G.~M\"unch suggested this project; I am greatly indebted
to him and to A.~Mampaso for their advice during its development.
I also thank F.~P.~Keenan and S.~N.~Nahar for sending their data in
electronic form, and M.~A.~Bautista for kindly providing the [\ion{Fe}{ii}]
emissivities. I am grateful to T.~Mahoney for revising the English text and to
J.~E.~Beckman and W.~Wall for some later corrections.
I thank two anonymous referees for useful comments that have helped to
improve this paper.
This research has made use of NASA's Astrophysics Data System Abstract Service.
A grant of the Spanish DGES PB97--1435--C02--01 provided partial support for
this work.
\end{acknowledgements}


\end{document}